\documentclass{llncs}

\usepackage[dvipdfmx]{graphicx}
\usepackage{amsmath}
\usepackage{amssymb}
\usepackage{bm}
\usepackage{theorem}
\usepackage{listings}
\usepackage{color}
\usepackage{url}
\usepackage{subfigure}
\usepackage{multirow}
\usepackage{bussproofs}
\usepackage[linesnumbered,ruled,vlined]{algorithm2e}
\usepackage{mathtools}
\usepackage{wrapfig}
\usepackage{article}

\usepackage{scalerel}
\newcommand\mathbox[1]{\mathord{\ThisStyle{%
  \fboxsep2\LMpt\relax\kern1\LMpt\fbox{$\SavedStyle#1$}\kern1\LMpt}}}

\begin{document}

\mainmatter

\title{Compositional Test Generation of\\ Industrial Synchronous Systems}

\author{Daisuke Ishii\inst{1} \and Takashi Tomita\inst{1} \and Kenji Onishi\inst{2} %\and Shigeki Takeuchi\inst{2} 
  \and Toshiaki Aoki\inst{1}}
\institute{Japan Advanced Institute of Science and Technology\\
  \email{\{dsksh,tomita,toshiaki\}@jaist.ac.jp} \and 
  Gaio Technology Co., LTD. ~
  \email{onishi.k@gaio.co.jp} }
  %\email{\{??,takeuchi.s\}@gaio.co.jp} }

\maketitle

\begin{abstract}
    Synchronous systems provide a basic model of embedded systems and 
    industrial systems are modeled as Simulink diagrams and/or Lustre programs.
    Although the test generation problem is critical in the development of safe systems,
    it often fails because of the spatial and temporal complexity of the system descriptions.
    This paper presents a compositional test generation method to address the complexity issue.
    We regard a test case as a counterexample in safety verification, and
    represent a test generation process as a deductive proof tree built with dedicated inference rules;
    we conduct both spatial- and temporal-compositional reasoning along with a modular system structure.
    %In a deduction, we utilize a classical assume-guarantee rule by Alur and Henzinger, and also a rule for temporal composition.
    A proof tree is generated using our semi-automated scheme involving manual effort on contract generation and automatic processes for counterexample search with SMT solvers.
    As case studies, the proposed method is applied to four industrial examples involving such features as enabled/triggered subsystems, multiple execution rates, filter components, and nested counters.
    In the experiments, we successfully generated test cases for target systems that were difficult to deal with using the existing tools.
\end{abstract}

\section{Introduction}
\label{s:intro}

Many industrial products containing embedded computers,
as exemplified typically by automobiles,
are developed using a model-based approach.
\emph{Synchronous systems} are useful in the co-modeling of such embedded systems~\cite{Alur2015b}.
In industrial sites, dedicated modeling tools %for modeling synchronous systems 
such as MATLAB/Simulink and SCADE are used to achieve rapid prototyping of the products.
Because the products are to be deployed in a safety-critical environment, assuring their quality is mandatory, as standards such as ISO~26262 demand.
A state-of-the-art solution to this is \emph{coverage testing}, which ensures that a product is exercised exhaustively to satisfy a criterion consisting of \emph{objective properties} in a system description.
In the modeling phase, developers invest effort in building a test suite that fulfills coverage criteria, e.g. MC/DC, with tool support. For example, Simulink is equipped with the \emph{Simulink Design Verifier} (SLDV) toolbox for this purpose.

Automated test-suite generation scales up to a certain complexity and product size;
however, we are aware of cases in which the testing phase becomes stuck as a result of uncovered objectives.
Test generation tends to fail because of system descriptions unsupported by the tools and the computational complexity of searching for a test case.
As for generic software, there have been a number of test generation methods proposed for synchronous systems.
For instance, it has been reported that an incomplete approach, e.g. a random search or a Monte-Carlo method, increases the coverage rate~\cite{Satpathy2008,Pasareanu2009,Kanade2009,Peranandam2012,Matinnejad2016a,Tomita2020}.
However, in our experiments with industrial products, there still remain some objectives not achieved through using the available tools.

Industrial synchronous systems involve both spatial and temporal complexity and any methods/tools may suffer from it.
\emph{Spatial complexity} is primarily introduced by a composition of multiple atomic nodes, which might be connected circularly.
\emph{Temporal complexity} is introduced in a system by nodes involving long counters and nodes that behave with different clocks.
Complexity is increased further by additional features such as composition of filter nodes and combinational nodes and equipment of an activation port to nodes.

%There is a dilemma between the rigorousness of quality assurance and the complexity of the problem of test generation.

While there might always be a complex system that is difficult to analyze with existing methods,
we take a compositional approach using a divide-and-conquer process on the spatial and temporal composition of a target.
Using a tool such as Kind2~\cite{Champion2016b}, which supports compositional description and model checking, can be expected to be effective.
The objective of this research is to leverage this idea in a semi-automated method and to apply it to several industrial systems.
The contributions of this paper are as follows:
\begin{itemize}
    \item \emph{A compositional test generation method based on static analysis with SMT solvers}.
        We first formalize the problem in terms of safety verification, constructing a counterexample compositionally.
        After describing the link between \emph{Simulink diagrams}, \emph{Lustre programs}, and a classical \emph{assume-guarantee (AG) rule} by Alur and Henzinger~\cite{Alur1999}, we propose inference rules for spatial and temporal composition.
        To reduce the complexity and to facilitate our compositional reasoning, we express test cases and interface contracts between modules with \emph{templates} of input/output streams.
        Finally, we propose a set of manual and automated methods to identify test cases and contracts.
    \item \emph{Case studies on compositional test generation of industrial examples}.
        First, we present two example Simulink models developed in an industrial site and extract their essential parts in two other simpler examples.
        We then report how the proposed method is applied to test generation of the examples.
        Last, we report the experimental results including comparisons with the results of two existing tools.
\end{itemize}

The remainder of the paper is organized as follows.
Sect.~\ref{s:related} reviews related work.
Sect.~\ref{s:sync} introduces basic definitions that formalize synchronous systems, a motivating example, and two languages (modeling tools), Simulink and Lustre.
Sect.~\ref{s:form} formalizes the test generation problem and provides templates for streams.
Sect.~\ref{s:comp} presents our method for compositional test generation.
In Sect.~\ref{s:ex}, we describe the four examples and the test generation process.
Sect.~\ref{s:xp} reports the experimental results.

\section{Related Work}
\label{s:related}

Proprietary tools such as 
%\emph{Simulink Design Verifier} (SLDV)
\emph{SLDV}\footnote{\url{https://www.mathworks.com/products/simulink-design-verifier.html}.}
and \emph{Reactis}\footnote{\url{https://www.reactive-systems.com/}.}
are often used
in the testing phase of industrial synchronous systems.
Although these produce test cases for achieving a high coverage rate for many systems, there are those whose descriptions are not supported and those for which test generation takes a long time.
The examples in this paper include such cases.

Various test generation methods for synchronous systems have been proposed, including
symbolic execution with random search~\cite{Satpathy2008,Pasareanu2009,Kanade2009,Peranandam2012},
Monte-Carlo numerical simulation~\cite{Tomita2020},
meta-heuristic search~\cite{Matinnejad2016a},
model-based testing~\cite{Gadkari2008}, and
mutation testing~\cite{Zhan2005,Garoche2014}.
%conformance testing~\cite{mmt2014}. %\cite{Abbas2014}.
Each of these analyzes the target in its entirety as a monolithic system and can encounter excessive amount of computation when a system becomes large and test cases become rare.
Testing methods have been proposed in other domains that combine compositional analysis and random search~\cite{Godefroid2007,Giannakopoulou2008,McMillan2019}.
We expect that such methods will be studied in the future for synchronous systems.
%
%\Todo{Falsification methods}

Compositional modeling and verification of synchronous systems have been studied for decades~\cite{Halbwachs1993,Alur1999,Giannakopoulou2018}.
This work aims to apply their results to industrial case studies, primarily by using the AG rule in \cite{Alur1999} for circular compositions.
Henzinger and others have also considered composition and abstraction on the temporal direction~\cite{Alur1999,Henzinger1999}. 
We propose a different rule for temporal composition.
More recently, compositional model checking for Lustre programs has been implemented in the tools, Kind2~\cite{Champion2016b,Champion2016} and Zustre,\footnote{\url{https://github.com/coco-team/zustre}.} which make use of state-of-the-art SMT solvers.
The AG reasoning by Kind2 addresses only the spatial direction of modules with a restricted form of composition.
This paper proposes a more generic spatial-temporal composition method using the rule in \cite{Alur1999} and other rules.
Some algorithms for compositional analysis on circular modules (which are not synchronous systems) have been proposed~\cite{AbdElkader2018,McMillan2019}.
Automation for synchronous systems is a future work.

%%%%

\section{Synchronous Systems}
\label{s:sync}

In this work, we consider synchronous systems taken from industry. %, which are typically created with commercial tools.
We describe a basic model in Sect.~\ref{s:sync:basic}, two dedicated languages, Simulink and Lustre, in Sect.~\ref{s:sync:lang}, some extensions for industrial use in Sect.~\ref{s:sync:ext}, and
the complexity of industrial systems in Sect.~\ref{s:sync:complexity}.

\subsection{Basic Synchronous Model}
\label{s:sync:basic}

We use \emph{typed variables} to describe synchronous systems.
Types include \texttt{bool}, \texttt{int} and \texttt{real}, referring to $\{\True,\False\}$, $\mathbb{Z}$ and $\mathbb{R}$, respectively.
We also consider \emph{typed expressions} constructed with typed variables.
%, logical operators and type-specific vocabularies..

As a basic model of discrete, reactive and synchronous computation, we consider a set of compositional nodes represented as Mealy machines~\cite{Halbwachs1993,Alur1999,Alur2015b}.

\begin{definition}
A \emph{node} of name $n$ is a tuple $N_n = (I_n, O_n, S_n, \Init_n, \AB \React_n)$ where
each of $I_n$, $O_n$ and $S_n$ is a set of input/output/state variables ($I_n \cap O_n \cap S_n = \emptyset$), 
$\Init_n$ is an \emph{initial condition description}, and
$\React_n$ is a \emph{reaction description}.
We denote the domains of $I_n$, $O_n$ and $S_n$ by $\Dom{I_n}$, $\Dom{O_n}$ and $\Dom{S_n}$, respectively.
$\Init_n$ is interpreted as a subset of $\Dom{S_n}$.
We have a \emph{reaction} %$(s_j \xrightarrow{i_{j+1}/o_{j+1}} s_{j+1})$ 
in $\Dom{S_n}\Times\Dom{I_n}\Times\Dom{O_n}\Times\Dom{S_n}$
by an interpretation of $\mathit{React}_n$;
and an \emph{execution} of a node (of length $k$) is formalized as a \emph{stream} i.e. a sequence of reactions
\begin{equation*}
    s_{-1} \xrightarrow{i_0/o_0} s_0 \xrightarrow{i_1/o_1} s_1 \cdots s_{k-2} \xrightarrow{i_{k-1}/o_{k-1}} s_{k-1},
\end{equation*}
where $s_\Box \in \Dom{S_n}$, $i_\Box \in \Dom{I_n}$ and $o_\Box \in \Dom{O_n}$ and $s_{-1}$ satisfies $\Init_n$.
%
%We denote the set of executions of length $k$ by $\mathit{Execs}_k$.
A sequence of values $(i_0/o_0 \cdots i_{k-1}/o_{k-1})$ %in $\Dom{I_n}$ and $\Dom{O_n}$ 
taken from an execution is called a \emph{trace}.
\end{definition}

Next, we introduce the implementation relation and the composition mechanism for the synchronous nodes, following the formalization in \cite{Alur1999}.
\begin{definition}
We say a node $N_m$ \emph{implements} a node $N_n$ %that represents a safety property, 
if
(i)~$O_n \subseteq O_m$,
(ii)~$I_n \subseteq I_m \cup O_m$,
(iii)~a dependency in the evaluation of $x \in I_n \cup O_n$ on $y \in I_n$ is preserved in $N_m$, and
(iv)~for every trace $t$ of $N_m$, the projection of $t$ onto $O_n$ is a trace of $N_n$.
We denote this relation by $N_m \models N_n$.
%
%\Todo{We write $N_m \equiv N_n$ iff $N_m \models N_n$ and $N_n \models N_m$.}
\end{definition}
The implementation relation is reflexive and transitive.
\begin{definition}
Given two \emph{compatible} nodes $N_n = (I_n,O_n,S_n, \Init_n, \React_n)$, where $n \in \{1,2\}$, $O_1 \cap O_2 = \emptyset$ and $\React_1 \cup \React_2$ is acyclic,
we consider the \emph{parallel composition} $N_1 \PC N_2 = (I,O,S,\Init,\React)$, where $I = (I_1 \cup I_2) \setminus O$, $O = O_1 \cup O_2$, $S = S_1 \cup S_2$, $\Init = (\Init_1,\Init_2)$ and $\React = \React_1 \cup \React_2$.
\end{definition}
Commutativity and associativity of the composition operator follow from the definition.
The languages described in the next subsection support an equivalent of composite nodes, and their processors must check various properties, including acyclicity above and causality, based on the DAG representation of $\React_n$.
We allow implicit renaming of state variables if a collision occurs in a composition.
Otherwise, a renaming of a variable $x$ (of whatever kind) to $y$ is denoted by $N_n[x := y]$.
We also denote replacement of an initial condition $\Init_n$ with $e$ by $N_n[\Init_n := e]$.

\begin{example}
\label{s:sync:basic:ex}
    We model a counter with a ``hold'' mode as a synchronous node $N_\texttt{Cnt}$, where
    $I_\texttt{Cnt} = \{\texttt{En}\}$, $O_\texttt{Cnt} = \{\texttt{C}\}$, and $\mathit{React}_\texttt{Cnt}$ is described as
    \begin{equation*}
        \texttt{C} = 
        \begin{cases}
            %\texttt{C\char`_} 
            1 + \mathit{pre}(\texttt{C})
            & \text{if $\texttt{En} = \True$}, \\
            \mathit{pre}(\texttt{C}) & \text{otherwise}.
        \end{cases}
        %\qquad
        %\texttt{C\char`_} = 1 + \mathit{pre}(\texttt{C}).
    \end{equation*}
    Here (and hereafter), the expression $\mathit{pre}(X)$ is interpreted as ``the value of expression $X$ in the previous round,'' and we formalize it with a state variable named ``$\mathit{pre}(X)$.''
    Accordingly, the state variable set is to be $S_\texttt{Cnt} = \{\mathit{pre}(\texttt{C})\}$, and the initial condition is $\mathit{Init}_\texttt{Cnt} \equiv (\mathit{pre}(\texttt{C})=0)$.
    As an example of executions of $N_\texttt{Cnt}$, when an input stream $(\False\ \True\ \False\ \True)$ is fed, the output is $(0\ 1\ 1\ 2)$.
\end{example}

%%%%

\subsection{Modeling Languages}
\label{s:sync:lang}

We assume two representations of synchronous nodes, Simulink diagrams and Lustre programs, described with the respective languages.
Our case study systems are originally modeled with Simulink, and we use Lustre to explain the proposed method in the following sections.

\begin{figure}[!t]
    \centering
    \includegraphics[width=.95\textwidth]{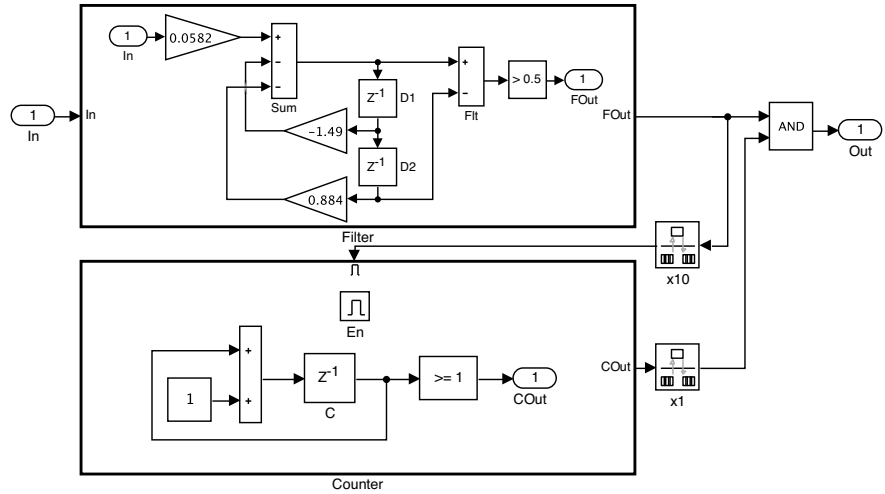} 
    \caption{Example Simulink model: ``A filter and a counter'' (\texttt{Sys1}).}
    \label{f:ex1}
\end{figure}

%\subsubsection{MATLAB/Simulink.}
\emph{Simulink}\footnote{\url{https://www.mathworks.com/products/simulink.html}} is a MATLAB toolbox for modeling synchronous (and hybrid) systems using a GUI. 
It can be used to model with either a continuous timeline or a timeline discretized with a fixed \emph{sample time}; in this work, we assume the latter and abstract each time with the number of accumulated sample times, as in \cite{Tripakis2005,Bourbouh2020}.
Simulink diagrams, often called \emph{models}, are hierarchical directed graphs with edges, called \emph{lines}, and nodes, called \emph{blocks}, of various kinds. Fig.~\ref{f:ex1} shows an example.
Simulink models are modularized by encapsulating portions into \emph{subsystem} blocks. The example in Fig.~\ref{f:ex1} contains two subsystems, \verb|Filter| and \verb|Counter|. 
In this work, we regard each Simulink model and each subsystem in a model as a synchronous node.
In addition, each line is regarded as a variable of a node.
Lines of Simulink are typed as machine-representable numbers e.g. \verb|int64| and \verb|double|. In this work, we idealize them as mathematical types; %e.g. with \verb|int| and \verb|real|; 
analysis with actual types is a future work.

%\subsubsection{Lustre.}
\emph{Lustre}~\cite{Caspi1987} is a programming language for describing synchronous nodes.
For example, the following code describes the content of the upper subsystem in Fig.~\ref{f:ex1} as a node.
\begin{lstlisting}
node Filter (In: real) returns (FOut: bool)
var Sum, D1, D2, Flt: real;
let
  Sum = 0.0582*In - (-1.49*D1) - 0.884*D2;
  D1 = 0.0 -> pre Sum; %\hspace{1em}% D2 = 0.0 -> pre D1;
  Flt = Sum - D2;      %\hspace{2em}% FOut = Flt > 0.5;
tel
\end{lstlisting}
A Lustre program is easily associated with a synchronous node counterpart.
The code above describes a node with the elements
$I_\texttt{F} = \{\texttt{In}\}$,
$O_\texttt{F} = \{\texttt{FOut}\}$,
%$S_\texttt{F} = \{\texttt{Sum}, \mathit{pre}(\texttt{Sum}), \texttt{D1}, \mathit{pre}(\texttt{D1}), \texttt{D2}, \texttt{Flt}\}$,
$S_\texttt{F} = \{\mathit{pre}(\texttt{Sum}), \mathit{pre}(\texttt{D1})\}$, and
$\Init_\texttt{F} \equiv %(\forall x \!\in\! S_\texttt{F}, x = 0)$ 
(\mathit{pre}(\texttt{Sum}) = \mathit{pre}(\texttt{D1}) = 0)$
(the name \verb|Filter| is abbreviated).
Note that, we need not regard \texttt{Sum}, \texttt{D1}, \texttt{D2} and \texttt{Flt} as variables because their values are local to an evaluation of the reaction at a round.
The \lstinline|let| section describes $\Init_\texttt{F}$ and $\React_\texttt{F}$.
For instance, the sentence \lstinline{D1 = 0.0 -> pre Sum;} is interpreted as follows.
In the first transition, \verb|D1| is assigned $0$ (the initial value of $\mathit{pre}(\texttt{Sum})$ is $0$);
in the following transitions, \verb|D1| is assigned the value of \verb|Sum| in the previous round.

There are several works on translation from Simulink to Lustre~\cite{Caspi2003,Tripakis2005,Bourbouh2020}.
We assume that Simulink models are translated using a variant of the existing methods.
Some of the translation techniques are described in the next subsection.

%

%\subsection{Example: A Filter and A Counter}

\begin{example}
\label{s:sync:lang:ex}
We consider ``a filter and a counter,'' a synchronous system that consists of a second-order filter and a counter. This is represented as the Simulink diagram in Fig.~\ref{f:ex1}.
In Lustre, the top-level system is described as
\begin{lstlisting}
node Sys1 (In: real) returns (Out: bool) var FOut, COut: bool;
let FOut = Filter(In); COut = Counter(RateTransition(FOut, 10));
%\hspace{2em}%Out = FOut and COut;%\hfill%tel
\end{lstlisting}
The description of \verb|Filter| is shown above.
\verb|RateTransition| is a predefined node %for a rate transition, 
and will be explained in Sect.~\ref{s:form:template}.
The \verb|Counter| node is described as
\begin{lstlisting}
node Counter (En: bool) returns (COut: bool) var En_: bool; C_,C: int;
let En_ = En -> ((pre En_) or En);
%\hspace{2em}%C_ = 0 -> if (not pre En_) then 0 else (1 + pre C);
%\hspace{2em}%C = 0 -> if En then C_ else pre C;
%\hspace{2em}%COut = false -> if En then C >= 1 else pre COut;%\hfill%tel
\end{lstlisting}
The description of the reaction is more complicated than Ex.~\ref{s:sync:basic:ex}, and the behavior is slightly different.
Some details of the description and ideas behind it are explained in the following.
\end{example}

\subsection{Other Features of Industrial Simulink Models}
\label{s:sync:ext}

This section describes major features found in industrial Simulink models, and how they are translated into Lustre.

The first feature we examine is \emph{conditionally executed subsystems} (CESs). These are equipped with 
%an auxiliary input port 
\verb|Enable| and \verb|Trigger| ports
to control whether the subsystem is activated.
In Fig.~\ref{f:ex1}, the \verb|Counter| subsystem equips an \verb|Enable| port (named \verb|En|).
The subsystem updates the output value iff the enable signal is non-zero; otherwise, the previous output value is retained unchanged (by default).

We translate a CES into a Lustre node with additional input variables that represent the control; every sentence in its description is conditioned with the control inputs.
In \cite{Tripakis2005}, CESs are described in a similar manner but using Lustre's \verb|when| operator, which is not supported in some implementations (e.g. Kind2).
An example is the description of the \verb|Counter| node.
For each reaction, its original description, e.g. the assignment to \verb|C_|, and the conditioning with the control input, e.g. the assignment to \verb|C|, are described.
Special care is taken for cases in which the node is disabled initially and later outputs the initial value $0$ when activated.

Second, industrial models often involve \emph{multiple rates}, the use of several sample times.%
\footnote{Simulink also handles variable \emph{sample offsets}. We omit this for the sake of simplicity.}
%
%A Simulink model is specified a sample time as a parameter of the model or a block.
%%(it is set implicitly with the GUI).
%When multiple sample times are used, they are propagated through the model; 
%some blocks e.g. \verb|Subsystem| assume that inputs of the same rate are fed and an inconsistency results in the failure of a simulation; 
%other blocks e.g. \verb|Switch| resolves an inconsistency of the input rates with the \emph{GCD rule} (refer for details to \cite{Tripakis2005}).
%%
%Sample times are also specified with the \verb|RateTransition| block, which specifies the output sample time.
%
Sample times are specified in a model in various ways. 
In Fig.~\ref{f:ex1}, two \verb|RateTransition| blocks specify the sample time of their output streams.
The block \verb|x1| sets the ``default'' sample time to $1$;
then, for the \verb|Counter| subsystem, the sample time is set to $10$ with another block \verb|x10|.

Although translation from a multi-rate Simulink model into Lustre can be performed simply with the \verb|when| operator~\cite{Tripakis2005,Bourbouh2020}, 
we translate such models in a lengthy way (without \verb|when|) as follows:
(i) we determine the default sample time as the GCD of the specified sample times;
(ii) when multiple sample times are used in a subsystem, we describe each portion with the same sample time as a Lustre node, and describe the subsystem as a composite node;
(iii) when a subsystem (or a portion) has a coarser sample time than the default, it is translated into a Lustre node with an additional input variable, as for the \verb|Enable| port;
(iv) a Lustre node with a slower rate is instantiated with a \verb|RateTransition| node of the specified rate given as an input.
An example translation is shown as the \verb|Sys1| and \verb|Counter| nodes, where the input \verb|En| of \verb|Counter| is also used for the clock control and is fed \verb|RateTransition(FOut, 10)| in \verb|Sys1|.

Third, Simulink supports handling \emph{multiplex streams} by grouping several lines with the \verb|Mux| block.
These can be simply translated into Lustre using Lustre's arrays or records~\cite{Tripakis2005}.
In this work, we translated them using a set of bare variables to avoid possible overhead.

\subsection{Spatial and Temporal Complexity of Synchronous Systems}
\label{s:sync:complexity}

\emph{Spatial complexity} correlates with the number of node compositions made.
Conversely, large system descriptions tend to involve multiple synchronous nodes as in Ex.~\ref{s:sync:lang:ex}.
Then, analysis of a composite description can be processed node-wise.
However, the connection between the nodes matters in a compositional analysis;
for instance, circular referencing between two nodes and CESs that introduce dependency between nodes make it difficult to analyze separately.

The features in previous subsection may affect also the \emph{temporal complexity}.
Use of CESs brings complexity not only in space but also in time because a CES operates on activated rounds, which can be segmented.
Multiple rates in a system increase the number of (global) rounds taken in a sub-node due to the rate ratio.
Also, filter nodes such as \verb|Filter| in Ex.~\ref{s:sync:lang:ex}, well characterized in the frequency domain, %are used with combinational nodes such as conditional branching on states.a 
bring a difficulty to analyze them in the time domain.

%%%%

\section{Test Generation and Template Nodes}
\label{s:form}

In this work, test generation is regarded as safety verification, which checks whether the target top-level node satisfies the test objective in a round of an execution (Sect.~\ref{s:form:problem}).
In Sect.~\ref{s:form:template}, we introduce templates of synchronous nodes to specify test cases and intermediate signal streams that serve as {contracts} in compositional reasoning (cf. Sect.~\ref{s:comp}).

\subsection{Test Generation Problem}
\label{s:form:problem}

We consider \emph{safety properties} $\phi$ on a synchronous node $N_n$ to describe test objectives and intermediate properties.
\begin{definition}
    The syntax of \emph{safety properties} $\phi$ is as follows.
    \begin{align*}
        e &~::=~ \text{(Predicate on $O_n$)} ~|~ \text{(Boolean combination of $e$)}, \\
        \phi &~::=~~ e ~~|~~ \At{e}{s} ~~|~~ \phi \land \phi,
    \end{align*}
    where $s \in \mathbb{N}$ represents a round of executions.
\end{definition}
Intuitively, for every trace of $N_n$, a property of the form $e$ (resp. $\At{e}{s}$) means that the output in every round (resp. in the round $s$) satisfies $e$.
We denote by $o \models \phi$ that the valuation with an output value $o \in \Dom{O_n}$ satisfies $\phi$.

\begin{definition}
    We interpret a safety property $\phi$ as a synchronous node $N_{\PN{\phi}} = (\emptyset,O_\PN{\phi},S_\PN{\phi},\Init_\PN{\phi},\React_\PN{\phi})$ with nondeterministic behaviors.
    When $\phi \equiv e$, $S_\PN{\phi} = \emptyset$ and
    $\React_\PN{\phi}$ describes $\{s_\emptyset \xrightarrow{/o} s_\emptyset ~|~ o \models e \}$;
    when $\phi \equiv \At{e}{s}$, $S_\PN{\phi} = \{c\}$, $\Init_\PN{\phi} = 0$ and $\React_\PN{\phi}$ describes $\{s_c \xrightarrow{/o} s_c+1 ~|~ (s_c,o) \models (c = s \Rightarrow e) \}$;
    when $\phi \equiv \phi_1 \land \phi_2$, we rename the counter variables if necessary, and merge the reaction descriptions of $N_\PN{\phi_1}$ and $N_\PN{\phi_2}$.
\end{definition}
A node $N_{[\phi1 \land \phi2]}$ is equivalent to $N_{[\phi1]} \PC N_{[\phi2]}$ iff $O_{[\phi1]} \cup O_{[\phi2]} = \emptyset$.

\begin{definition}
    Given a \emph{target node} $N_n$ and an \emph{objective} %$\exists s, \At{e}{s}$,
$\Obj$ (a proposition on $O_{n}$),
    a \emph{test generation problem} asks to find a \emph{test case} $T$, a synchronous node, such that $T \PC N_n ~\models~ N_\PN{\At{\Obj}{s}}$ holds for some $s$.
\end{definition}

\subsection{Synchronous Node Templates}
\label{s:form:template}

To synthesize test cases and intermediate properties in a test generation, we use \emph{templates} to represent a set of streams of interest.
We can consider Lustre's \lstinline|const|-value streams as a template for constant-value streams.
Other templates are described using synchronous nodes with constant parameters and input streams;
for example, the $\texttt{Step}$ and $\texttt{Square}$ templates are described as follows.
\begin{lstlisting}
node Step (const s: int; v1, v2: 'a) returns (Out: 'a)
var c: int;
let c = 0 -> (pre c) + 1; Out = if c < s then v1 else v2;%\hfill%tel

node Square (const t, p: int; v1, v2: 'a) returns (Out: 'a) 
var c, l: int;
let c = 0 -> (pre c) + 1;
%\hspace{2em}%l = 0 -> if c+p-(pre l) >= 2*t then (pre l) + 2*t else (pre l);
%\hspace{2em}%Out = if c+p-l >= t then v1 else v2;%\hfill%tel
\end{lstlisting}
These nodes represent respectively a temporal concatenation of two streams \verb|v1| and \verb|v2| at the round \verb|s| and a square wave of half-length \verb|t| and phase shifting \verb|p| that evaluates to either \verb|v1| or \verb|v2|.
In the descriptions above, \texttt{'a} denotes a polymorphic type, which is used for space efficiency and is to be replaced with a primitive type regarding the context.
The last two inputs of \verb|Step| and \verb|Square| are not \lstinline|const| for utility (e.g. to build multiple step streams).
In addition, we restrict the parameter values to $\texttt{s} > 0$ for \texttt{Step} and $\texttt{t} \geq 2$ and $0 \leq \texttt{p} < 4 \texttt{t}$ for \texttt{Square}.\footnote{These constraints are specified as preconditions using Kind2's annotation language.}

\begin{example}
\label{s:form:ex1}
    A test objective for checking that the \verb|Filter| node in Fig.~\ref{f:ex1} outputs the value $\True$ at step 10 is described by a safety property $\At{(\texttt{FOut} = \True)}{10}$ (or simply $\At{\texttt{FOut}}{10}$).
    The node implements a band-path filter for which the \verb|Flt| block outputs greater values when a signal stream with a specific frequency is input.
    For instance, the objective is fulfilled with the input $\texttt{Square}(5,1,-1,1)$.
\end{example}

As an adapter between nodes with different rates, we use the template \verb|RateTransition| for nodes that synchronize a stream \verb|En_| with a clock signal stream of interval \verb|s|.
\begin{lstlisting}
node RateTransition (En_: bool; const s: int) returns (En: bool)
var c: int;
let c = 1 -> if pre c >= s then 1 else (pre c) + 1;
%\hspace{2em}%En = if c <= 1 then En_ else false;%\hfill%tel
\end{lstlisting}
Every node that runs with a slower rate than the default is equipped with a control input (see Sect.~\ref{s:sync:ext}); hence, a \verb|RateTransition| instance is set at the upstream to configure the rate.
%\footnote{As exemplified in Ex.~\ref{s:form:ex2}, a composition with \texttt{RateTransition} requires% a careful renaming of the variable names of the control inputs.}
In the following, we denote a \verb|RateTransition| instance with interval $s$ connected to a node $N_n$ by $\mathit{RT}_{s,n}$.

%\subsection{Example}

\begin{example}
\label{s:form:ex2}
    A composition of \verb|Filter| and \verb|Counter|, as in Fig.~\ref{f:ex1}, must be adapted with a \verb|RateTransition| instance to modify the rate of \verb|Counter|.
    The composition is described as
    \begin{equation*}
        N_\texttt{Filter} \,||\,
        N_\texttt{RateTransition}[\texttt{En\char`_} := \texttt{FOut}, \texttt{s} := 10] \,||\,
        N_\texttt{Counter}.
    \end{equation*}
    We abbreviate the above as $N_\texttt{Filter} \,||\, \mathit{RT}_{10,\texttt{Counter}} \,||\, N_\texttt{Counter}$.
\end{example}

%%%%

\section{Compositional Test Generation}
\label{s:comp}

In this section, we present a compositional test generation scheme, summarized in the following four steps.
\begin{enumerate}
    \item \emph{Specification of the composition}.
        Our method assumes that modellers specify a systems as a composition of synchronous nodes.
        Along with the composition, we construct an outline proof tree based on the dedicated inference rules (Sect.~\ref{s:comp:rules}).
        Here, the nodes of the tree are declarations of implementation relations, in which some nodes in the composites might be parameterized.
    \item \emph{Inference of test cases and contracts}.
        Each of the declarations above can contain a special node that represents a test case or a \emph{contract}, an intermediate property on the node interface. We leave this unknown in Step~1; in this step, we fill the blanks using both automated inference performed by SMT solvers and additional human efforts.
    \item \emph{Consistency checking between contracts}.
        Once the parameters are instantiated, the tree is viewed as an outline of compositional safety verification. 
        Some parts of it are validated automatically using existing tools (e.g. Kind2).
        If validation succeeds, the test cases generated in Step~2 can be expected to work for the entire system.
    \item \emph{Refinement based on a counterexample}.
        Failure of the check in Step~3 results in a counterexample, which is an execution that violates a contract.
        After analyzing it, we can go either to Step~2 to modify the contracts or 
        to Step~1 to modify the deduction tree e.g. by applying a temporal decomposition rule.
\end{enumerate}

\subsection{Translation into Combinational Nodes}
\label{s:comp:combi}

\emph{Combinational nodes}~\cite{Alur2015b} are synchronous nodes with no state variables.
For the purpose of temporal-compositional analysis of nodes, we translate a node into a combinational one in such a way that the internal states become externally visible and their initial values are given externally.
\begin{definition} \label{def:combi}
    Let $N_n$ be a node and $e_{Sn}$ be a proposition on $S_n$. 
    $C(N_n, e_{Sn})$ represents a composition of the following two nodes:
    \begin{enumerate}
    \renewcommand{\labelenumi}{(\roman{enumi})}
    \item The (combinational) node $(I_n\!\cup\!S_n, O_n\!\cup\!S'_n, \AB \emptyset, \True, \AB \React'_n)$, where $S'_n$ is a set of fresh variables $\{x' ~|~ x \in S_n\}$, and $\React'_n$ is equivalent to $\React_n$ except that the assignments 
    $x' := \mathit{next}(x)$ are appended for each $x \in S_n$,
    where $\mathit{next}(x)$ represents the value of $x$ in the next round, which must be described in $\React_n$.
    \item 
    %The safety property node $N_\PN{\At{e}{0}}$.
    The node $(S'_n, S_n, S''_n, e''_{Sn}, \React''_n)$, where $S''_n = \{\mathit{pre}(x') ~|~ x' \in S'_n\}$,
    $e''_{Sn}$ is equivalent to $e_{Sn}$ except that $x \in S_n$ is replaced with $x'' \in S''_n$,
    and $\React''_n$ describes the assignments $\forall x\!\in\!S_n, \AB x := \mathit{pre}(x')$.
    \end{enumerate}
    %
    %$C(N_n, e)$ represents the node $(I_n \cup S_n, O', \emptyset, e, \AB \React')$, where $O'$ is a set such that $O_n \subseteq O$ and, for each $x \in S_n$, it contains a fresh variable $x_n$, and $\React'$ is equivalent to $\React_n$ except that the assignments of the values of $x \in S_n$, which are described in $\React_n$, to the new variables $x_n$ are appended.
\end{definition}

\begin{example}
    Let us consider the node \texttt{Cnt} that is a translation of Ex.~\ref{s:sync:basic:ex} in Lustre.
    \begin{lstlisting}
node Cnt (En: bool) returns (C: int)
let C = if En then 1 + pre C else pre C;%\hfill%tel
    \end{lstlisting}
    Here, we denote the state variable $\mathit{pre}(\texttt{C})$ by \texttt{pC}.
    The two components of $C(N_\texttt{Cnt}, \AB \Init_\texttt{Cnt})$ are described as follows.
    \begin{lstlisting}
node Cnt_c1 (En: bool; pC: int) returns (C, pC': int)
let C = if En then 1 + pC else pC;
%\hspace{2em}%pC' = C;%\hfill%tel

node Cnt_c2 (pC': int) returns (pC: int)
let pC = 0 -> pre pC'%\hfill%tel
    \end{lstlisting}
\end{example}

\begin{lemma}
    For a node $N_n$, (i) $C(N_n, \Init_n) \models N_n$, and
    (ii) for every trace $t$ of $N_n$, there is a trace of $C(N_n, \Init_n)$ whose projection onto $O_n$ is $t$.
\end{lemma}

\subsection{Deduction System for Compositional Reasoning}
\label{s:comp:rules}

We propose a deduction system %set of inference rules in Fig.~\ref{f:rules} 
for constructing a proof tree for compositional reasoning regarding the implementation relation between synchronous nodes.

\begin{definition} \label{def:rules}
    The rules AG, Temp, RT, Cons and IP are as shown in Fig.~\ref{f:rules}.
\end{definition}

\begin{figure}[t]
\fbox{
\begin{minipage}{.96\textwidth}
    \begin{minipage}{.5\textwidth}
        \begin{prooftree}
            \AxiomC{$N_1 \,||\, N_b ~\models~ N_a$}
            \noLine
            \def\extraVskip{1pt}
            \UnaryInfC{$N_2 \,||\, N_a ~\models~ N_b$}
            \def\extraVskip{2pt}
            \RightLabel{\scriptsize(AG)}
            \UnaryInfC{$N_1 \,||\, N_2 ~\models~ N_a \,||\, N_b$}
        \end{prooftree}
    \end{minipage}
%    \begin{minipage}{.5\textwidth}
%        \begin{prooftree}
%            \AxiomC{$N_n ~\models~ N_{[\phi1]}$}
%            \noLine
%            \def\extraVskip{1pt}
%            \UnaryInfC{$N_n ~\models~ N_{[\phi2]}$}
%            \def\extraVskip{2pt}
%            %
%            \RightLabel{\scriptsize(LAnd)}
%            \UnaryInfC{$N_n ~\models~ N_{[\phi1 \land \phi2]}$}
%        \end{prooftree}
%    \end{minipage}
%    \begin{minipage}{.5\textwidth}
%        \begin{prooftree}
%            \AxiomC{$N_1 \,||\, N_b ~\models~ N_a$}
%            \AxiomC{$N_2 \,||\, N_a ~\models~ N_b$}
%            %
%            \RightLabel{\scriptsize(AG)}
%            \BinaryInfC{$N_1 \,||\, N_2 ~\models~ N_a \,||\, N_b$}
%        \end{prooftree}
%    \end{minipage}
    %
    \begin{minipage}{.5\textwidth}
        %\vspace*{.5em}
        \begin{prooftree}
            \AxiomC{$C(N_n, \Init_n) ~\models~ N_\PN{\At{e'_{Sn}}{j}}$}
            \noLine
            \def\extraVskip{0pt}
            \UnaryInfC{$C(N_n, e_{Sn}) ~\models~ N_\PN{\At{e}{k}}$}
            \def\extraVskip{2pt}
            \RightLabel{\scriptsize(Temp)}
            \UnaryInfC{$N_n ~\models~ N_\PN{\At{e}{j+k+1}}$}
        \end{prooftree}
    \end{minipage}
    \begin{minipage}{\textwidth}
        \vspace*{.5em}
        \begin{prooftree}
            %\AxiomC{$\vphantom{C\models}$}
            %\noLine
            %\def\extraVskip{1pt}
            \AxiomC{$N_n \PC \mathit{RT}_{s,n} ~\models~ N_\PN{\phi}$}
            \RightLabel{\scriptsize(RT)}
            \UnaryInfC{$N_n[@\Box := @r\Box] \PC \mathit{RT}_{rs,n}
            ~\models~ N_\PN{\phi}[@\Box := @r\Box]
            $}
        \end{prooftree}
    \end{minipage}
%    \begin{minipage}{.6\textwidth}
%        \vspace*{.5em}
%        \begin{prooftree}
%            \AxiomC{$C(N_n, \Init_n) ~\models~ N_\PN{\At{e'_{Sn}}{j}}$}
%            \noLine
%            \def\extraVskip{1pt}
%            \UnaryInfC{$C(N_n, e_{Sn}) ~\models~ N_\PN{\At{e}{k}}$}
%            \def\extraVskip{2pt}
%            %
%            \RightLabel{\scriptsize(Temp)}
%            \UnaryInfC{$N_n ~\models~ N_\PN{\At{e}{j+k+1}}$}
%        \end{prooftree}
%    \end{minipage}
%    %\vspace*{-1.5em}
%    \begin{minipage}{.38\textwidth}
%        \vspace*{.5em}
%        \begin{prooftree}
%            \AxiomC{$N_a ~\models~ N_b$}
%            \noLine
%            \def\extraVskip{1pt}
%            \UnaryInfC{$N_1 \models N_a$ \quad $N_b \models N_2$}
%            \def\extraVskip{2pt}
%            %
%            %\insertBetweenHyps{\hskip 5pt}
%            \RightLabel{\scriptsize(Cons)}
%            \UnaryInfC{$N_1 ~\models~ N_2$}
%        \end{prooftree}
%    \end{minipage}
    %
    \begin{minipage}{.6\textwidth}
        \vspace*{.5em}
        \begin{prooftree}
            \AxiomC{$N_1\ \models\ N_a$\hspace{-1em}}
            \AxiomC{$N_a\ \models\ N_b$\hspace{-1em}}
            \AxiomC{$N_b\ \models\ N_2$}
            \insertBetweenHyps{\hskip 8pt}
            \RightLabel{\scriptsize(Cons)}
            \TrinaryInfC{$N_1 ~\models~ N_2$}
        \end{prooftree}
    \end{minipage}
    \begin{minipage}{.4\textwidth}
        \vspace*{.5em}
        \begin{prooftree}
            \AxiomC{$\vphantom{N_1 ~\models~ N_2}$}
            \RightLabel{\scriptsize(IP:$\phi'\!\Rightarrow\!\phi$)}
            \UnaryInfC{$N_\PN{\phi'} ~\models~ N_\PN{\phi}$}
        \end{prooftree}
    \end{minipage}
\end{minipage}
}
    \caption{Inference rules for compositional test generation. $N_\Box$, where $\Box \in \{1,2,a,b,n\}$, represents an arbitrary synchronous node.}
    \label{f:rules}
\end{figure}

Compositional reasoning for systems that involve synchronized concurrent nodes is conducted using \emph{assume-guarantee} (AG) methods~\cite{Giannakopoulou2018,Alur1999},
with each node in the system associated with a contract, an assumption on the inputs, and a guarantee on the outputs.
In Fig.~\ref{f:rules}, the rule AG is intended for the AG reasoning on spatial-compositional nodes.
%where $N_1$, $N_2$, $N_a$ and $N_b$ are synchronous nodes.
This rule allows each component $N_1$ and $N_2$ to be analyzed separately when combined with a portion of the consequence nodes.
Given a safety property $N_{[\phi1 \land \phi2]}$, it is translated as $N_{[\phi1]} \PC N_{[\phi2]}$ (when $O_{[\phi1]} \cup O_{[\phi2]} = \emptyset$) and an AG composition can be applied.

The rule Temp provides a method for temporal-compositional reasoning.
When verifying a property at the $(j\!+\!k\!+\!1)$-th round, it splits the reachability problem into two portions and allows verification with analyzing up to round $j$ or $k$.
To observe the internal states and modify the initial states, this rule requires the transformation in Def.~\ref{def:combi}.
Let $e$, $e_{Sn}$ and $e'_{Sn} $be propositions on the variables in $O_n$, $S_n$ and $S'_n$ (cf. Def.~\ref{def:combi}), respectively.
%$\mathit{Ext}(N_n, e_S)$ is a modified node of $N_n$ of which every internal state in $S_n$ is made externally visible, i.e. $O_n$ and $S_n$ are modified as $O_n := O_n \cup S_n$ and $S_n := \emptyset$; also, its initial condition is replaced with $e_S$.

The rule RT enables temporal shrinking of a verification task to $1/r$ ($r \in \mathbb{N}_{>1}$) when a rate transition is applied to the target node.
$N_n[@\Box := @r\Box]$ denotes to replace every property of the form $e@\Box$ described in $N_n$ to $e@r\Box$.
With this rule, an analysis up to step $k$ suffices when checking $\phi \equiv \At{e'}{rk}$.

The rule Cons represents logical consequence with respect to the implementation relation.
The rule IP translates logical consequence between contracts into the implementation relation.

\begin{lemma}
    The rules in Def.~\ref{def:rules} (and Fig.~\ref{f:rules}) are sound.
\end{lemma}
\emph{Proof sketch.}
The soundness of AG is proved in \cite{Alur1999}.
For Temp, suppose that the goal does not hold and contains a counterexample.
Then, if the state in the $j$-th round of the counterexample does not satisfy $e_{Sn}$, it contradicts with the first premise;
otherwise, it contradicts with the second premise.
For other rules, the soundness is straightforward from the definitions of $RT_{rs,n}$, implementation relation, and safety properties.

\subsection{Basic Procedures}
\label{s:comp:procs}

This section proposes some procedures for Steps~2--4 for performing static analyses on node descriptions.
%Including automated methods and strategies for manual efforts

\subsubsection{Model checking using an SMT solver.}
Given a node $N_n$ and a safety $\At{e}{k}$, \emph{bounded model checking} (BMC)~\cite{Biere1999} checks whether $N_n \models N_\PN{\At{e}{k}}$ using an off-the-shelf SMT solver.
In our compositional deduction, we rely on this procedure; in addition to the rules in Fig.~\ref{f:rules}, we use the rule below to represent a proof made with a BMC.
\begin{prooftree}
    \AxiomC{}
    \RightLabel{\scriptsize(V)}
    \UnaryInfC{$N_n ~\models~ N_\PN{\At{e}{k}}$}
\end{prooftree}
A BMC process is based on an encoding of the fact $N_n \models N_\PN{\neg e}$ into a logic formula.
If checking the satisfiability by a trace of length $k+1$ results in UNSAT, then the fact $N_n \models N_\PN{\At{e}{k}}$ is proved.
Otherwise, a result of SAT indicates a {counterexample}, a valuation of the encoded formula that represents an execution satisfying $\At{(\neg e)}{k}$.
%
%A BMC process
%(i) encodes the executions of length $k$ and the property $\neg \At{e}{k}$ into logic formulas $f_{N_n,k}$ and $f_{\neg e,k}$, respectively; and
%(ii) checks the satisfiability of $f_{N_n,k} \land f_{\neg e,k}$ with an SMT solver.
%When the result is UNSAT, the fact $N_n \models P_{\At{e}{k}}$ is proved.
%When the result is SAT, we have a \emph{counterexample}, which is an assignment to the variables of $f_{N_n,k}$ and representing a trace that satisfies $\neg e$ at step $k$.
%
%In the following experiments, we use Kind2 that implements the BMC on Lustre programs.

\subsubsection{Test generation as safety falsification.}
Given a test generation problem with an objective $\Obj$, we use a BMC to generate a test case $T$ automatically.
Here, in contrast to the previous paragraph, we perform a BMC to falsify the safety $\neg \Obj$; hence, we encode $\At{\Obj}{k}$, for each $k \geq 0$.
A result of SAT for some $k$ indicates a counterexample for $\neg \Obj$; then, we construct a test case $T$ such that $T \PC N_n \models N_\PN{\At{\Obj}{k}}$ holds. %(it will be confirmed with another BMC).

%To prove efficiently, it is useful to perform the following induction.
%\begin{prooftree}
%    \AxiomC{
%    \begin{tabular}{ll}
%        (Base case) &
%        $N ~\models~ P_I \Rightarrow 
%        %\Evt_{[1]}\ 
%        p_O$ \\
%        (Induction step) &
%        $N^* ~\models~ P_I \Rightarrow \Alw p_S$ \qquad
%        $N[\mathit{Init} \mapsto p_S] ~\models~ p_O$
%    \end{tabular}
%    }
%    \UnaryInfC{$N ~\models~ P_I \Rightarrow \Alw p_O$}
%\end{prooftree}
%The base case is proved by a BMC on $N \models \Alw_{[1..k]}\ \neg p_O$.
%We can introduce an overapproximation $p_S$ with e.g. a numerical reachability analysis and a propagation of the constants described in $N$.
%Note that the induction is not sound since $p_S$ is an overapproximation of the intermediate states.

\subsubsection{Test generation with templates.}
In general, test generation considers a large search space i.e. arbitrary streams of length $k+1$ or less.
We are able to use the template nodes in Sect.~\ref{s:form:template}
to reduce the search space to a set of meaningful signal streams, depending on the context. 
We perform a BMC on the safety of the system $T' \PC N_n$ using a template $T'$;
then, it instantiates the template parameters with a SAT valuation.

\subsubsection{Pre- and postcondition inference.}

As mentioned earlier, contract inference is important in compositional reasoning.
Basically, we deduce contracts manually based on the reaction description of a node.
Otherwise, the following methods will be useful:
(i)~\emph{Constant propagation}. A system description may contain a constant, e.g. a threshold used for a case analysis. It is useful for bounding input/output domains.
(ii)~\emph{Invariant generation}. Recent SMT solvers are equipped with a function for deducing a lemma from a given formula. This function is useful for deducing invariants of synchronous nodes; e.g., Kind2 implements this function.
(iii)~\emph{Use of templates.} We can assume a template for an interface of a composite and try to generate an invariant for its parameters.

%%%%

\section{Examples}
\label{s:ex}

\begin{wraptable}[5]{r}{0.5\textwidth}
\centering
    \vspace{-2em}
\begin{tabular}{l|rrrr}
    \hline
    & Sys.~1 & Sys.~2 & Sys.~3 & Sys.~4 \\
    \hline
    $|I|$/$|O|$ & $1/1$ & $2/1$ & $4/4$  & $19/2$ \\
    \# blocks   & $22$  & $54$  & $215$  & $281$ \\
    \# compo's  & $3/0$ & $3/2$ & $4/14$ & $3/4$ \\
    \hline
\end{tabular}
\end{wraptable}
This section describes four examples of compositional test generation following the steps explained at the beginning of the previous section.
The latter two target systems, Sys.~3 and Sys.~4, are excerpted from industrial products whose test objectives are difficult to achieve using existing tools, as explained later.
The first two systems, Sys.~1 and Sys.~2, are simplified versions of Sys.~3 and Sys.~4, respectively, intended to explain the compositional process using the proposed method.
Each example is described with both Simulink and Lustre.
Some data on the Simulink models representing their sizes are shown in the table above, which shows
the numbers of the top-level input variables $|I|$ and output variables $|O|$,
the number of blocks in the Simulink model (excluding the inport/outport blocks), and
the number of components in the direction of space/time handled by the proposed method.
At the end of every case study, the test case generated is confirmed to achieve the objective with simulation using Simulink.

\subsection{A Filter and a Counter}
\label{s:ex:1}

We recall Ex.~\ref{s:sync:lang:ex}.
The objective here is to observe the output (\verb|Out|) becomes $\True$ in a bounded execution, i.e. $\Obj \equiv (\texttt{Out} = \True)$.
%
%We first present a proof tree representing the compositional reasoning on that a given test case fulfills the objective at a step.
%Then, we explain the building process of the tree, which involves identification of several parameters, e.g., the test case, the step when the objective is fulfilled, and intermediate properties to connect the deductions.

For the composition of the back-end part, we modify the node \verb|Sys1| as:
\begin{lstlisting}
node Sys1_m (In: real) returns (Out, FOut, COut: bool)
let FOut = Filter(In); COut = Counter(RT(FOut, 10)); 
%\hspace{2em}%Out = Back(FOut, COut);%\hfill%tel
\end{lstlisting}
Here, the node \verb|Back| is described as:
\begin{lstlisting}
node Back (FOut, COut: bool) returns (Out: bool)
let Out = FOut and COut;%\hfill%tel
\end{lstlisting}

\begin{figure}[t]
\fbox{
    \begin{minipage}{.96\textwidth}
    \vspace{-1em}
    \begin{align}
        \label{eq:ex1:f}
        \mathbox{T} \PC N_\texttt{F} \PC \mathbox{N_\PN{\At{\texttt{COut}}{20}}} &~\models~ \mathbox{N_\PN{\At{\texttt{FOut}}{10} \land \At{\texttt{FOut}}{20}}} \\
        \label{eq:ex1:c}
        \mathit{RT}_{1,\texttt{C}} \PC N_\texttt{C} \PC \mathbox{N_\PN{\At{\texttt{FOut}}{1} \land \At{\texttt{FOut}}{2}}}  &~\models~ \mathbox{N_\PN{\At{\texttt{COut}}{2}}} \\
        \label{eq:ex1:fc}
        \mathbox{T} \PC N_\texttt{F} \PC \mathit{RT}_{10,\texttt{C}} \PC N_\texttt{C} &~\models~ \mathbox{N_\PN{\At{\At{\texttt{FOut}}{10} \land (\texttt{FOut} \land \texttt{COut})}{20}}} \\
        \label{eq:ex1:fc:out}
        \mathbox{T} \PC N_\texttt{F} \PC \mathit{RT}_{10,\texttt{C}} \PC N_\texttt{C} \PC N_\PN{\At{\texttt{Out}}{\mathbox{20}}} &~\models~ \mathbox{N_\PN{\At{(\texttt{FOut} \land \texttt{COut})}{20}}} \\
        \label{eq:ex1:b}
        N_\texttt{B} \PC \mathbox{N_\PN{\At{(\texttt{FOut} \land \texttt{COut})}{20}}} &~\models~ N_\PN{\At{\texttt{Out}}{\mathbox{20}}} \\
        \label{eq:ex1:goal}
        \mathbox{T} \PC N_\texttt{F} \PC \mathit{RT}_{10,\texttt{C}} \PC N_\texttt{C} \PC N_\texttt{B} &~\models~ N_\PN{\At{\texttt{Out}}{\mathbox{20}}}
    \end{align}
    \begin{prooftree}
        \AxiomC{}
        %\RightLabel{\scriptsize (V, Cons)}
        \RightLabel{\scriptsize (V)}
        \UnaryInfC{\eqref{eq:ex1:f}}
        \AxiomC{}
        \RightLabel{\scriptsize (V)}
        \UnaryInfC{\eqref{eq:ex1:c}}
        %
        %\RightLabel{\scriptsize (RT)}
        %\UnaryInfC{\eqref{eq:ex1:c:rt}}
        %
        \RightLabel{\scriptsize (RT,AG)}
        \BinaryInfC{\eqref{eq:ex1:fc}}
        \RightLabel{\scriptsize (Cons,IP)}
        \UnaryInfC{\eqref{eq:ex1:fc:out}}
        \AxiomC{}
        \RightLabel{\scriptsize (V)}
        \UnaryInfC{\eqref{eq:ex1:b}}
        \RightLabel{\scriptsize (AG,Cons)}
        \BinaryInfC{\eqref{eq:ex1:goal}}
    \end{prooftree}
    \end{minipage}
}
    \caption{Outline of test generation of ``a filter and a counter'' (\texttt{Sys1\char`_m}).}
    \label{f:ex1:tree}
\end{figure}

As for Step~1, we determine an outline of the test generation process based on the composition of the nodes \verb|Front|, \verb|Counter| and \verb|Back| (henceforth, \verb|F|, \verb|C| and \verb|B|, respectively),
which is illustrated as the proof tree in Fig.~\ref{f:ex1:tree}.
The tree consists of the declarations of the implementation relations \eqref{eq:ex1:f}--\eqref{eq:ex1:goal}.
The terms surrounded by boxes are unknown in Step~1 and are identified in the later steps.
In the tree, the test case is denoted by $T$. 
As a result, the proof succeeds with 
\begin{equation} \label{eq:ex1:tc}
    T ~=~ N_{\texttt{Square}(5,1,-1,1)}[\texttt{Out}:=\texttt{In}],
\end{equation}
which is an instance of the \texttt{Square} template placed at the input of $N_\texttt{F}$.

Let us look into the tree from bottom to top.
At the bottom, the AG rule is applied, with the system divided as $T \PC N_\texttt{F} \PC RT_{10,\texttt{F}} \PC N_\texttt{C}$ and $N_\texttt{B}$. Before applying AG, we strengthen the consequent by a composition with a contract node; the contract $\At{(\texttt{FOut} \land \texttt{COut})}{20}$, identified in Step~2, is a precondition for $N_\texttt{B}$ to output $\True$ (it is checked by Eq.~\eqref{eq:ex1:b}).
In the other branch, Eq.~\eqref{eq:ex1:fc:out} is obtained by strengthening the antecedent and weakening the consequent of Eq.~\eqref{eq:ex1:fc}.
Above that, the AG rule is applied again with the composition of $T \PC N_\texttt{F}$ and $RT_{10,\texttt{F}} \PC N_\texttt{C}$.
To obtain Eq.~\eqref{eq:ex1:c}, we apply the RT rule to shrink the timeline for $1/10$.
The consequent $\At{\texttt{COut}}{2}$ of Eq.~\eqref{eq:ex1:c} does not affect the other branch (Eq.~\eqref{eq:ex1:f}), because the scopes do not overlap, i.e., $\texttt{COut} \notin O_\texttt{T} \cup O_\texttt{F}$.
%$\texttt{Out} \notin \mathit{V}(N_\texttt{F} \PC N_\texttt{C})$

The inference of the contracts (boxed terms in Fig.~\ref{f:ex1:tree}) performed in Step~2 is outlined as follows.
\begin{enumerate}
\renewcommand{\theenumi}{\roman{enumi}}
    \item We obtain the property $\exists s, \At{(\texttt{FOut} \land \texttt{COut})}{s}$ by a precondition calculus for the postcondition $\exists s, \At{\texttt{Out}}{s}$ of $N_\texttt{B}$.
    \item We analyze $N_\texttt{C}$ using an SMT solver.
        We assume that the constant enable signal $\True$ is provided and falsify the property $\neg \texttt{COut}$ (we omit $\neg \texttt{FOut}$ since \verb|FOut| is out of the scope).
        It succeeds with $k = 2$.
    \item We check whether a stream of length 21 whose values are $\True$ in rounds $10$ and $20$ can be output from $N_\texttt{F}$.
        For this check, we add to $N_\texttt{F}$ a monitor %for the output before 10~rounds,
        %which is described in $N_\texttt{F}$ as 
\begin{lstlisting}[basicstyle=\ttfamily\footnotesize]
  FOut_d10 = Delay <<10>> (false, FOut);
\end{lstlisting}
%\begin{lstlisting}[basicstyle=\ttfamily\footnotesize]
%  FOut_d10 = %$\overbrace{\texttt{false -> pre(}}^{10}$% FOut %$\overbrace{\texttt{)}}^{10}$%
%\end{lstlisting}
%\begin{lstlisting}[basicstyle=\ttfamily\footnotesize]
%  FOut_d10 = false -> pre(false -> pre(false -> pre(false -> pre(
%    false -> pre(false -> pre(false -> pre(false -> pre(
%    false -> pre(false -> pre( FOut ))))))))));
%\end{lstlisting}
        using a variable delay node whose output is \verb|false| initially and \verb|FOut| delayed for 10~rounds in the following.
        Then, we falsify $\neg (\texttt{FOut\_d10} \land \texttt{FOut})$,
        assuming a square-wave signal stream.
        The falsification does not succeed with $k = 10$ but succeeds with $k = 20$, yielding the test case~\eqref{eq:ex1:tc}. This proves Eq.~\eqref{eq:ex1:f}.
    \item Finally, we assume the enable stream for the rounds 1 and 2 and falsify $\neg \At{\texttt{COut}}{2}$. It fails, thereby proving Eq.~\eqref{eq:ex1:c}.
\end{enumerate}

\subsection{Two Guarded Decrementers}
\label{s:ex:2}

\begin{figure}[!t]
    \centering
    \includegraphics[width=\textwidth]{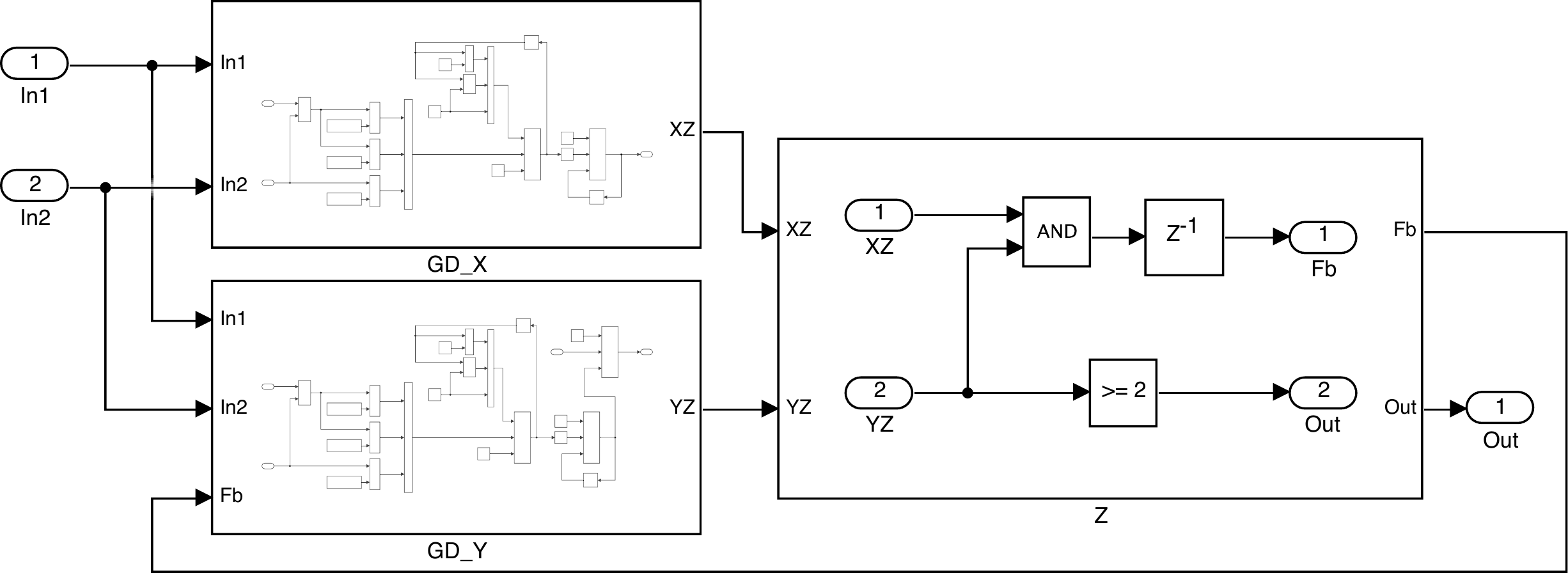} 
    \caption{Simulink model of ``two guarded decrementers'' (\texttt{Sys2}).}
    \label{f:ex2}
\end{figure}

The second example involves a circular interconnection between components and also illustrates the use of the rule Temp for temporal decomposition.
Fig.~\ref{f:ex2} shows a Simulink model of the example, which consists of three nodes (subsystems). 
The model receives two real-valued streams as inputs and outputs a Boolean stream.
The nodes \verb|GD_X| and \verb|GD_Y| are almost equivalent except for the condition described and an additional conditioning at the end of \verb|GD_Y|.
Each of them describes a control gate for checking some conditions on the input values, and also a decrementer that affects the decision by the gate; we omit the details owing to space limitations.
%\Todo{Example of a behavior.}
Another node \verb|Z| receives inputs \verb|XZ| and \verb|YZ| of types \verb|bool| and \verb|int| and checks some conditions;
the value of \verb|XZ| represents whether decrementing finishes ($\True$) or not ($\False$);
\verb|YZ| evaluates to either $2$ (when \verb|Fb| is true), $1$ (when decrementing finishes) or $0$ (otherwise).

We consider the test generation problem with $\Obj \equiv (\verb|Out| = \True)$.
We translate the Simulink model into a Lustre program and apply the proposed method.
As in the previous subsection, we show the resulting proof tree and then explain the process of compositional test generation.

\begin{figure}[t]
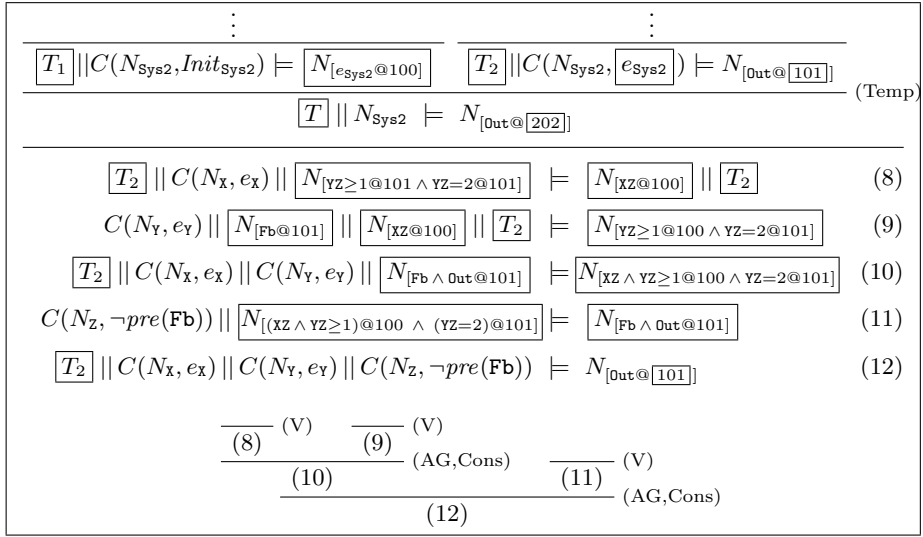

\fbox{
    \begin{minipage}{.96\textwidth}
    \vspace*{-1em}
    \begin{prooftree}
        \AxiomC{\vdots}
        %\RightLabel{\scriptsize (Pre)}
        \UnaryInfC{$\mathbox{\vphantom{N_[}T_1} \!\PC\! C(N_\texttt{Sys2},\!\Init_\texttt{Sys2}) \models \mathbox{N_\PN{\At{e_\texttt{Sys2}}{100}}}$}
        \AxiomC{\vdots}
        %\RightLabel{\scriptsize (Pre)}
        \UnaryInfC{\hspace{-1pt}$\mathbox{\vphantom{N_[}T_2} \!\PC\! C(N_\texttt{Sys2},\!\mathbox{\vphantom{N_[}e_\texttt{Sys2}}) \models N_\PN{\At{\texttt{Out}}{\mathbox{101}}}$}
        \insertBetweenHyps{\hskip 5pt}
        \RightLabel{\scriptsize (Temp)}
        \BinaryInfC{$\mathbox{T} \PC N_\texttt{Sys2} ~\models~ N_\PN{\At{\texttt{Out}}{\mathbox{202}}}$}
    \end{prooftree}
    \vspace{-.5em}
    \hrule width 1\textwidth
    \begin{align}
        \label{eq:ex2:x}
        \mathbox{T_2} \PC C(N_\texttt{X},e_\texttt{X}) \PC \mathbox{N_\PN{\At{\texttt{YZ}\geq1}{101} \,\land\, \At{\texttt{YZ}=2}{101}}} ~&\models~ \mathbox{N_\PN{\At{\texttt{XZ}}{100}}} \PC \mathbox{T_2} \\
        \label{eq:ex2:y}
        C(N_\texttt{Y},e_\texttt{Y}) \PC \mathbox{N_\PN{\At{\texttt{Fb}}{101}}} \PC \mathbox{N_\PN{\At{\texttt{XZ}}{100}}} \PC \mathbox{T_2} ~&\models~ \mathbox{N_\PN{\At{\texttt{YZ}\geq1}{100} \,\land\, \At{\texttt{YZ}=2}{101}}} \\
        \label{eq:ex2:xy}
        \mathbox{T_2} \PC C(N_\texttt{X},e_\texttt{X}) \PC C(N_\texttt{Y},e_\texttt{Y}) \PC \mathbox{N_\PN{\At{\texttt{Fb} \,\land\, \texttt{Out}}{101}}} ~&\models\! \mathbox{\!N_\PN{\At{\texttt{XZ} \,\land\, \texttt{YZ}\geq1}{100} \,\land\, \At{\texttt{YZ}=2}{101}}\!} \\
        \label{eq:ex2:z}
        C(N_\texttt{Z},\neg\mathit{pre}(\mathtt{Fb})) \PC \mathbox{\!N_\PN{\At{(\texttt{XZ} \,\land\, \texttt{YZ} \geq 1)}{100} \ \land\ \At{(\texttt{YZ} = 2)}{101}}\!} \!&\models~ \mathbox{N_\PN{\At{\texttt{Fb} \,\land\, \texttt{Out}}{101}}} \\
        \label{eq:ex2:step2}
        \mathbox{T_2} \PC C(N_\texttt{X},e_\texttt{X}) \PC C(N_\texttt{Y},e_\texttt{Y}) \PC C(N_\texttt{Z},\neg\mathit{pre}(\mathtt{Fb})) ~&\models~ N_\PN{\At{\texttt{Out}}{\mathbox{101}}}
    \end{align}
    \vspace{-1.5em}
    \begin{prooftree}
        \AxiomC{}
        \RightLabel{\scriptsize (V)}
        \UnaryInfC{\eqref{eq:ex2:x}}
        \AxiomC{}
        \RightLabel{\scriptsize (V)}
        \UnaryInfC{\eqref{eq:ex2:y}}
        \RightLabel{\scriptsize (AG,Cons)}
        \BinaryInfC{\eqref{eq:ex2:xy}}
        \AxiomC{}
        \RightLabel{\scriptsize (V)}
        \UnaryInfC{\eqref{eq:ex2:z}}
        \RightLabel{\scriptsize (AG,Cons)}
        \BinaryInfC{\eqref{eq:ex2:step2}}
    \end{prooftree}
    \end{minipage}
}
    \caption{Outline of test generation of  ``two guarded decrementers'' (\texttt{Sys2}).}
    \label{f:ex2:tree}
\end{figure}

Fig.~\ref{f:ex2:tree} illustrates the proof tree.
The root of the overall tree is depicted in the top part.
We denote the overall composite node by $N_\texttt{Sys2}$ and the composite of the sub-nodes translated into combinational nodes (Def.~\ref{def:combi}) by $C(N_\texttt{Sys2},e)$, where $e$ is a conjunction of the initial conditions of the sub-nodes.
Again, the terms obtained by additional inferences are enclosed with boxes.
The test case $T$ is temporally decomposed into $T_1$ and $T_2$.
The generated $T$ in the end is composed of nested \verb|Step| templates, where the stream \verb|In1| is specified as a four-step stream
\begin{equation*}
T_\texttt{In1} ~=~ N_{\mathtt{Step}(101,\ \mathtt{Step}(1, 1.2, 0.8),\ \mathtt{Step}(176, -0.6, -0.8))},
\end{equation*}
where $T_\texttt{In1}$ represents a sub-node for the output \verb|In1|,
% T \equiv T_\texttt{In1} \PC T_\texttt{In1}
and the stream \verb|In2| is set as a two-step stream.

Fig.~\ref{f:ex2:tree} also illustrates the right-hand side branch that is built above the second premise of the root (we abbreviate \verb|GD_X| and \verb|GD_Y| as \verb|X| and \verb|Y|).
The sub-goal is discharged by a temporal decomposition.
It is based on a strategy of executing two decrementers in sequence (first \verb|X| and then \verb|Y|);
we verify the decrementing behavior of each node separately, for rounds $0,\ldots, 100$ and rounds $101,\ldots,202$.
$e_\texttt{X}$ and $e_\texttt{Y}$ represent the intermediate state at the round $100$.
The stream for the variable \verb|Fb| is expected to be valued $\False$ except for the final round $202$; thus, we set the initial condition $\neg\mathit{pre}(\mathtt{Fb})$ in Eq.~\eqref{eq:ex2:step2}.
The rest of the proof tree is constructed with an application of AG, some rewritings of properties, and node decompositions.

The inference of the contracts in Step~2 is outlined as follows:
\begin{enumerate}
\renewcommand{\theenumi}{\roman{enumi}}
    \item As a first trial, we attempt to falsify the objective without the temporal decomposition.
        We deduce a precondition $\exists s, (\texttt{XZ} \land \texttt{YZ}\geq 1)@s$ by analyzing the node \verb|Z|.
        Then, we can falsify the precondition for the nodes \verb|X| and \verb|Y|, both with the bound 101; the \verb|Square| template is set to \verb|In1| and \verb|Constant| is set to \verb|In2|;
        as a result, we obtain a test case $T_1$ or $T_2$ for \verb|X| or \verb|Y|, respectively.
        However, the two test cases differ and it can be verified with a further analysis that there are no common test cases of length 101 or less.
    \item We take the above-mentioned strategy to decrement \verb|X| and \verb|Y| in series and to decompose temporally.
        Applying the rule Temp requires the intermediate condition $e_\texttt{Sys2}$;
        here, we simply perform a numerical simulation to obtain the state at round $101$.
        For the first phase, we use $T_1$ as a test case and compute $e_\texttt{Sys2}$ with a simulation.
    \item For the second phase, by falsifying the precondition $\exists s, (\texttt{YZ}\geq 1)@s$ for the node $C(N_\texttt{Y},e_\texttt{Y})$, we obtain a test case $T_2'$ of length 101.
        We must check additionally that inputting $T_2'$ to $C(N_\texttt{X},e_\texttt{X})$ satisfies the precondition $\exists s, \texttt{XZ}@s$ after 101 rounds.
    \item Finally, we concatenate $T_1$ and $T_2'$ to obtain the test case $T$.
        To construct the proof tree, we also need contracts of $C(\texttt{Sys2},e_\texttt{Sys2})$ at round 100, i.e. $(\texttt{XZ} \land \texttt{YZ}\geq1)@100$, due to the feedback \verb|Fb|.
\end{enumerate}

\Todo{The round numbers should be double checked.}

\subsection{PM Motor Control}
\label{s:ex:3}

The third example, which describes a model of a motor and its controller, uses three sample times, $s=5\Times10^{-5}$, $20s$ and $100s$, and requires a large number of rounds to fulfill the objective at the base sample rate.
We generated a test case for observing a specific value as a top-level output.
We omit the details of the target because it is an in-house product and would be lengthy; instead, we report the process of compositional test generation.

We separated the top-level spatially into three modules and a sub-module into two; we denote the composition as $(N_1 \PC N_2) \PC (N_{20s} \PC N_{100s})$ (the latter names represent the local sample times).
In the system, the nested counter in the node $N_{20s}$ is the main cause of the test case being long.
Based on the numerical simulation of the system, we found that the objective requires decrementing the counter three times and it takes at least $865 \times 20 = 17\,300$ rounds.
To this end, we applied the rules RT and Temp to analyze the behavior in a temporally separated and shrunken way;
to separate the behavior into 12 segments of lengths $59$--$83$ as combinational nodes, we used the intermediate conditions on the internal counter that is incremented or reset every round.

The other nodes were exercised to synchronize with the behavior of $N_{20s}$.
Since the node $N_1$, which contains the output variable in the objective, is combinational, its invariant precondition for the objective was calculated.
$N_2$ was separated with Temp, and it was verified that the first module reaches a constant state with a particular input; then, the rate of the second module $N_{100s}$ was shrunk using RT, and it was verified that it fulfills a necessary intermediate condition induced from the objective with an instance of the \verb|Step| template of length $173$.

Finally, we obtained a test case of length $17\,301$, which consists of three one-step streams and one three-step stream.

\subsection{Extended Two Guarded Decrementers}
\label{s:ex:4}

The last example is an industrial model that involves two guarded decrementers as in \verb|Sys2|.
We report a summary of the test process.

We separated the model spatially into three as $N_1 \PC N_2 \PC N_3$;
$N_1$ is located upstream of others, and $N_2$ is located upstream of $N_3$.
Because we could not find, in the experiments, an intermediate condition (as a template instance) on the interface variables between $N_1$ and either of $N_2$ or $N_3$,
we analyzed $N_1 \PC N_2$ and $N_1 \PC N_3$ combined into one node each.

For the first composite $N_1 \PC N_2$, we performed a manual constant propagation in the node description, and we found that the node outputs a constant stream assuming some condition on the input.
This constant output was assumed as a contract in the analysis of the other composite.

The composite $N_1 \PC N_3$ contains the two decrementers and they are required to be exercised separately.
Different from the process in Sect.~\ref{s:ex:2}, we did not apply further spatial composition to $N_3$ because some peripheral behaviors were difficult to describe in a contract and the spatial size was not critical for analysis.
We simply applied the rule Temp to split the temporal analysis into four;
as a result, we obtained the fragments of the test case of length $50$ or $52$ for four combinational translations of $N_1 \PC N_3$.

The test case obtained is of length $204$, which consists of
(i) eight constant streams,
(ii) eight four-step streams,
(iii) two one-step streams, and
(iv) one composite square and four-step stream.

\section{Experiments}
\label{s:xp}

This section reports the experimental results on the test generation of Systems~1--4 in the previous section.
Sect.~\ref{s:xp:stat} shows the statistical data on the experiments using our compositional method.
Sect.~\ref{s:xp:compat} presents the results with the existing tools for comparisons.
The experiments were conducted using a 2.2GHz Intel Xeon E5-2650v4 processor with 128GB of RAM.
The descriptions in Lustre and Simulink and the SMT-LIB scripts for validation are available at {\url{https://dsksh.github.io/vmcai2021.zip}}.

\subsection{Compositional Test Generation}
\label{s:xp:stat}

We have partially automated the proposed method using Kind2 (version~1.2.0), which is equipped with the Z3 SMT solver (version~4.8.8).
In the experiments, first, most of the test cases and contracts were obtained by falsifying the objectives and intermediate conditions with BMC implemented in Kind2.
This is an incremental process in which the bounds to analyze are gradually increased; thus, a shortest counterexample will be obtained if one exists.
For comparison, we have also tried to generate test cases by falsifying the objectives against entire systems.
%
%Second, once the proof tree for compositional test generation was built, we described it as a Lustre program annotated with CoCoSpec~\cite{Champion2016};
%then, it was verified using Kind2 on the validity of the annotated contract (assumptions and guarantees) of each node and the consistency between the contracts for each composition;
%Here, the k-induction method is used for validating inferences with the V rule.
%% This validation can be omitted.
%The validation of some inferences was omitted since it was not supported by CoCoSpec,
%e.g., the inferences with AG on circular nodes in Sys.~2 and Sys.~3 and the inferences using Temp and RT.
%
Second, once the proof tree for compositional test generation was built, we described manually the inference rules in Def.~\ref{def:rules} and the tree in an SMT-LIB\footnote{\url{http://smtlib.cs.uiowa.edu}.} script;
then, it was processed by Z3 to validate every inference performed in the proof tree.
In the description, the rule RT was detailed to ease matching with the target terms for the efficiency of solving;
also, the rewriting rules for some equivalent terms were described.

\begin{table}[t]
    \centering
    \setlength{\tabcolsep}{5pt}
    \caption{\label{t:xp:stat} Experimental results.}
    \begin{tabular}{l|rrrr}
    \hline
    & Sys.~1 & Sys.~2 & Sys.~3 & Sys.~4 \\
    \hline
    Comp. & $(0.80+0.03)$s & $(12.2+0.09)$s & $(134+0.4)$s & $(250+13)$s \\
    Mono. &         $1.1$s &         $790$s & %$>3\,600$s 
        TO & TO \\
    SLDV  &      $0.03$s & Undecided & TO & TO \\
    TMC   & $65$s (112) & TO & $52$s (2) & TO \\
    \hline
    \end{tabular}
\end{table}

The CPU time taken in the test generation processes is shown in two of the rows of Table~\ref{t:xp:stat}.
The row ``Comp.'' shows the time in the form of $(x+y)$s, which means $x$s and $y$s were taken in the falsification using Kind2 and the validity/consistency checking using Z3, respectively.
The row ``Mono.'' shows the time taken in the monolithic falsification on the entire systems.
``TO'' is for the timeout in $3\,600$s.

\subsection{Comparison with Existing Tools}
\label{s:xp:compat}

Some of the examples have difficulties with the existing tools.
We give comparison results as evidence.

We first compared with the testing of Simulink models using SLDV (Sect.~\ref{s:related}) on MATLAB (R2019b).
%\emph{Simulink Design Verifier} (SLDV)%
%\footnote{\url{https://www.mathworks.com/products/simulink-design-verifier.html}.}
In the experiment, we specified $\neg \Obj$ as a ``Proof Objective'' in the Simulink model and performed a ``Property Proving'' with SLDV.
The results are shown in the row ``SLDV'' of Table.~\ref{t:xp:stat}.
For Sys.~1, a test case was generated more efficiently than in the analysis with Kind2;
however, for other examples, test generations failed or did not terminate in $3\,600$s; the execution for Sys.~2 reported ``UNDECIDED WITH COUNTEREXAMPLE'' after 14s, probably, in our view,  because some blocks used were not supported.

Second, we compared with an in-house tool that implements the \emph{template-based Monte-Carlo method}~\cite{Tomita2020} (called ``TMC'' here).
TMC repeatedly instantiates a given template for each input port of the Simulink model and simulates it until the objective is fulfilled.
We configured the templates that were identified in the previous section for each system;
the lengths of simulations were set ten times longer than the length of our test cases for the efficiency of TMC.
The row ``TMC'' shows the execution time and the number of simulations (the averages for ten executions).
A test case was generated for Sys.~1 but the result indicated that handling of filters was not efficient even with an appropriate template.
Sys.~3 was processed efficiently with one to four trials. In fact, the solutions of the objective are not rare in this example; in such a case, exercising with random input of a particular length is an efficient strategy.
n the contrary, test generation for Sys. 2 and 4 did not succeed with $33\,331$ and $6\,196$ trials in $5$ hours.

In the experiments, SLDV and TMC could not handle Systems~2--4 and Sys.~2 and 4, although they worked efficiently when possible.
We conjecture that SLDV uses static analysis on models and therefore suffers from the complexity issue.
When it fails, we cannot know the exact reason since it is a black-box tool.
Yet TMC is a simple method that considers the target as a black box, but it works efficiently for many systems~\cite{Tomita2020}.
However, since it is an incomplete method, there would be some instances that are difficult to handle.
For example, when an objective has a rare solution as in Sys.~2 and 4, TMC is disadvantageous.
% Template was given with our analysis ...
% In the experiments above, we limited the simulation length ...

%

\section{Conclusion and Future Work}
\label{s:concl}

We presented a compositional test generation method for synchronous systems and four case studies on industrial Simulink models.

Simulink models were translated into Lustre programs, and associated with classical formalizations of synchronous systems.
Test generation was performed using the proposed scheme, by first drawing a proof tree with ``blank squares''
and then filling the blanks using either a manual or an automatic process for inferring the contracts.
In test generation, the rules AG, Temp and RT address the spatial and temporal complexity, respectively.
It was shown that the method can be automated partially using Kind2 and Z3.
Full automation of the generation and validation of proof trees is left for future work.

In the case studies, it was necessary to understand the content of the system descriptions and to infer the contracts.
It could become routine that modelers provide such contracts preliminarily.
Otherwise, automated inference of the contracts and/or an interactive tool that encourages the modelers to provide hints for a contract could be developed in future work.
We also consider that specifying a contract will facilitate co-modeling as it will refine the compositional model of a system.
Our experiments imply that systems such as Sys.~4 can be analyzed using a random simulation.
Hybrid methods of compositional reasoning and random simulation might be another research direction.

\bibliographystyle{splncs03}
\bibliography{compositional}

\end{document}